\documentclass[aps,eqsecnum,amsmath,twocolumn]{revtex4}
\usepackage{graphics, setspace,epsfig,color,subfigure}
\usepackage[letterpaper, dvips,width=7.5in,height=8.5in,includemp=false]{geometry}
\usepackage[vcentermath]{}

\newcommand{\be}{\begin{equation}}
\newcommand{\ee}{\end{equation}}

\begin{document}

\title[title]{The Skyrmion strikes back:  baryons and a new large $N_c$ limit}
\author{Aleksey Cherman and Thomas D. Cohen}

\affiliation{Department of Physics \\
University of Maryland\\College Park, MD 20742}


\begin{abstract}
In the large $N_c$ limit of QCD, baryons can be modeled as solitons, for instance, as Skyrmions. This modeling has been
justified by Witten's demonstration that all properties of
baryons and mesons scale with $N_c^{-1/2}$ in the same way as the
analogous meson-based soliton model scales with a generic
meson-meson coupling constant $g$. An alternative large $N_c$
limit (the orientifold large $N_c$ limit) has recently been
proposed in which quarks transform in the two-index antisymmetric
representation of $SU(N_c)$.  By carrying out the analog of
Witten's analysis for the new orientifold large $N_c$ limit, we
show that baryons and solitons can also be identified in the
orientifold large $N_c$ limit. However, in the orientifold large
$N_c$ limit, the interaction amplitudes and matrix elements scale
with $N_c^{-1}$ in the same way as soliton models scale with the
generic meson coupling constant $g$ rather than as $N_c^{-1/2}$ as in the
traditional large $N_c$ limit.
\end{abstract}

\maketitle

\section{Introduction}
In 1973 't Hooft proposed a large $N_c$ limit for QCD
\cite{'tHooft} that has proved to be a powerful tool in studying
QCD and other strongly coupled gauge theories.  't Hooft's idea
was to generalize the gauge group of QCD from $SU(3)$ to
$SU(N_c)$, and take $N_c \rightarrow \infty$ while keeping $g^2
N_c$ and the number of flavors $N_f$ fixed.  In this limit quark
loops are suppressed, and non-planar diagrams are suppressed by a
factor of $N_c^{-2}$ for each handle.  This greatly reduces the
number of diagrams one must consider and allows one to make many
qualitative predictions.  For instance, quark-loop suppression
implies the OZI rule, and baryons can be treated as solitons in
the large $N_c$ limit
\cite{WittenOrig,WittenCurrent,WittenGlobal}. While this helps
explain an important qualitative feature of hadronic physics, it
does pose a phenomenological difficulty in relating the large
$N_c$ limit to the physical world of $N_c=3$.  To wit, there are the
important cases in which the OZI rule is badly violated, and they are not explained in the large $N_c$ limit. These
cases include the situations in which the $U(1)_A$ anomaly plays a
critical role, such as in the $\eta' - \pi$ mass difference.

A new large $N_c$ limit for QCD that was proposed by Armoni, Shifman, and
Veneziano \cite{ArmoniPRL,Armoni03,Armoni04,Armoni05} has received
considerable recent attention.  This limit, which they have
dubbed the `orientifold large $N_c$ limit', starts from the
observation that at $N_c=3$ a quark can be described in two
equivalent ways.  It can be described as a Dirac spinor field
transforming according to the fundamental representation of color
$SU(3)$ or, equivalently, as a Dirac spinor field transforming
according to the two-index anti-symmetric representation of color
$SU(3)$.   One can take a large $N_c$ limit starting from either
one of these two possibilities.  Starting from the fundamental
representation yields the 't Hooft (or, if one wishes,
``traditional'') large $N_c$ limit (TLNC limit), while using the
anti-symmetric representation yields the  new orientifold (or
``other'') large $N_c$ limit (OLNC limit).

The OLNC limit has a number of attractive features from a
theoretical perspective. It is inspired by and related to
supersymmetric Yang-Mills theory, and for one flavor allows one to
apply some of the powerful analytic tools and results of
supersymmetric Yang-Mills theory to QCD.  However, it is
important to note that the OLNC has important differences from
the TLNC.  While non-planar diagrams are suppressed in the OLNC
limit (similarly to the TLNC limit), quark loops are not
suppressed in the OLNC limit, since they, like gluons, carry two
color indices. This alters the nature of the large $N_c$ scaling
in the theory.  Most significantly it implies that an n-meson
vertex scales with $N_c$ differently in the two expansions:
\begin{eqnarray}
\Gamma_n & \sim & N_c^{2-n} \; \; \; \; \; \, {\rm (OLNC)} \nonumber \\
\Gamma_n & \sim & N_c^{1-n/2} \; \;  \; {\rm (TLNC)}\, .
\label{meson-scale}
\end{eqnarray}
These scaling relations show that in the OLNC limit, mesons behave analogously to glueballs in the TLNC limit\cite{Veneziano}.  This is as one would expect, since in the OLNC limit both quarks and gluons carry two color indices.

Apart from the above difference in the scaling of meson interactions, there is another important
distinction between the OLNC and the TLNC limit.  Since quark loops are
not suppressed in the OLNC limit, unlike the TLNC limit it does
not impose the OZI rule. This has the disadvantage of not
explaining a generic feature of hadronic phenomenology (that the
TLNC limit explains quite neatly).  However, it has the
compensating virtue of not requiring large $1/N_c$ corrections in
those situations where quark loops are important, such as in the
$\eta'-\pi$ mass difference.

Witten \cite{WittenOrig,WittenGlobal} showed that it is
natural to make an identification between baryons and solitons,
such as the Skyrmion, in the TLNC  limit. The evidence for this
was based on explicit calculations of the scaling of the baryon
and meson masses and scattering amplitudes with $N_c$.  It was
seen that \emph{all} properties of baryons and mesons scale with
$N_c^{-1/2}$ in the same way as an analogous meson-based
soliton model scales with a generic meson-meson coupling constant
$g$. It is important to determine
whether this baryon-soliton identification can be made in the new
OLNC limit.

At first sight it appears that the identification does not work:
the mass of baryons is usually thought to scale as $N_c^1$, while as pointed out by Armoni and Shifman\cite{ArmoniShifman},
the mass of Skyrmions in the OLNC limit scales as
$N_c^2$, creating an apparent contradiction. It is
not hard to see that the Skyrmion mass scales as $N_c^2$.  For illustration
consider the simplest Skyrmion for two massless
flavors. The Lagrangian density is given by
\be
    \mathcal{L}_{S}=\frac{f_{\pi}^2}{4} \mathrm{Tr}(L_{\mu} L^{\mu})
    + \frac{\epsilon^2}{4} \mathrm{Tr}([L_{\mu},L_{\nu}]^2) \, ,
    \label{SK}
\ee
 where the left chiral current $L_\mu$ is given by
$L_{\mu} \equiv U^{\dagger}
\partial_\mu U$, with $U \in SU(2)_f$ \cite{Skyrme,ANW}.  The $U$ field can be
written as $U = \exp \left ( i \vec{\tau} \cdot \vec{\pi}/f_\pi
\right )$, where $\vec{\pi}$ is the pion field.  Upon expanding the
pion field in the Lagrangian one sees that the n-meson vertices
agree with the generic scaling rules of Eq.~(\ref{meson-scale})
only if
\begin{eqnarray}
\epsilon \sim  N_c^{1/2} \; \; \;  f_\pi
\sim  N_c^{1/2}  \; \; \; &{\rm (TLNC) }&\nonumber \\ \epsilon
\sim N_c^{1} \; \; \; \;  \; \; \;  f_\pi \sim N_c^{1} \; .  \; \;
\; & {\rm (OLNC)}&
\label{epsscale}
\end{eqnarray}
 The mass of the
Skyrmion depends only on the parameters $f_\pi$ and $\epsilon$; the
standard variational treatment \cite{ANW} yields a soliton mass
given by $M_s = \overline{m} \epsilon f_\pi$ where $\overline{m}$
is a dimensionless number obtained from the solution of the
variational equation.  From the scaling behavior of $f_\pi$ and
$\epsilon$ in Eq.~(\ref{epsscale}), one sees that the soliton mass
scales as $M_s \sim N_c^2$.  Moreover, it is quite easy to see
that the scaling of the soliton mass with $N_c^2$ is generic; it
does not depend on the details of the particular Skyrmion
Lagrangian used.

However, Bolognesi \cite{Bolognesi} has shown that the discrepancy between a
soliton mass scaling as $N_c^2$ and a baryon mass scaling as
$N_c^1$ is due to a naive (and incorrect) expectation about the scaling of the
baryon mass.  In fact, Bolognesi showed that a color singlet baryon state in the OLNC
limit must contain at least $N_c(N_c-1)/2 \sim N_c^2$ quarks
\cite{Bolognesi}.  This suggests that baryon masses should scale
as $N_c^2$, not $N_c^1$, which eliminates the apparent
inconsistency.

Bolognesi's observation that order $N_c^2$ quarks are required to make
a baryon is clearly of paramount importance in the identification
of baryons as Skyrmions in the OLNC limit. Moreover, ref. \cite{Bolognesi} notes that the coefficient of the Wess-Zumino-Witten term must be $N_c(N_c-1)/2$, as one would expect in order for the identification to be consistent.  However, by itself this
is not sufficient. Recall that Witten's identification of baryons
as solitons in the TLNC limit required far more than the simple
observation that a baryon had at least $N_c^1$ quarks. Rather it
was based on the observations that
\begin{enumerate}
    \item The total contribution
to the mass of the baryon---including the energy of interaction
between the quarks via (multiple) gluon exchange---is of order
$N_c^1$;
    \item The characteristic $N_c$ scaling of all other observables
of baryons and mesons (such as scattering amplitudes or form
factors) is analogous to the scaling of the same quantities in
soliton models, provided one scales $g$, the characteristic
coupling in the soliton model, as $N_c^{-1/2}$.
\end{enumerate}
 In fact, these conditions were not demonstrated rigorously in ref.~\cite{WittenOrig}.
Rather, it was shown that 1) various typical classes of gluon
exchange diagrams contributing to the mass scaled as $N_c$
(counting the combinatoric factors) and 2) characteristic classes
of diagrams associated with the various observables scaled
appropriately once combinatoric factors were included.

The question addressed in this paper is whether hadronic properties in the OLNC limit have the same $N_c$ scaling as the properties of solitons, with the characteristic coupling constant $g$ in the soliton model scaling as $g \sim N_c^{-1}$.  Such a scaling rule is consistent both with the baryon
scaling as $N_c^2$, and with the meson-meson scattering amplitudes
given in Eq.~(\ref{meson-scale}).  It is not clear how to
demonstrate this in a completely rigorous manner.  However, a
demonstration with a degree of rigor comparable to Witten's
original analysis for the TLNC limit will presumably suffice to
make a compelling case. The goal of this paper is to provide such
a demonstration via the consideration of classes of diagrams in a
manner analogous to ref.~\cite{WittenOrig}.  This would essentially
complete the program begun in ref.~\cite{Bolognesi} of
establishing a Skyrmionic description of baryons in the OLNC
limit.

If one follows the arguments in this paper, it will be obvious
that all of the qualitative conclusions for scaling rules with
$N_c$ apply  equally to the case in which the quarks are taken to
be in the two-index symmetric representation. However, we focus
on the anti-symmetric case since it corresponds to the physical
world at $N_c=3$; the symmetric case does not.

The generalization of Witten's analysis to baryons in the
OLNC limit is not completely trivial; there is an important
subtlety for baryons in the OLNC limit which is not present in the
TLNC limit. The nature of the issue can be seen by looking at the
one-gluon exchange contribution to the baryon energy.  For the
TLNC limit, Witten showed these contributions scale as $N_c^1$
(ref. \cite{WittenOrig}).   In a representative diagram of two
quarks interacting via a single gluon exchange, there are two
gluon vertices which together contribute a factor of $1/N_c$, and
a combinatoric factor of $N_c^2$ since each end of the gluon can
connect to one of the $N_c$ distinct quarks in the baryon.

A naive  generalization of this reasoning to the OLNC limit
suggests that there the one-gluon exchange contribution to the
mass scales like $N_c^3$.  There is again a $1/N_c$ factor for
the gluon vertices, but in the OLNC limit case there are
$N_c(N_c-1)/2 \sim N_c^2$ species of quark and thus the
combinatoric factor appears to scale as $N_c^4$.  If the
contribution of the one-gluon exchange contribution to the
nucleon mass really does scale as $N_c^3$, it suggests that the
baryon mass grows with $N_c$ faster than $N_c^2$, apparently
preventing an identification of baryons with the Skyrmions in the
OLNC limit.

In this paper, we demonstrate that despite the apparent
discrepancy above, the one-gluon exchange contribution to the
baryon mass scales only as $N_c^2$. As will be seen, there is an
important difference in the nature of one-gluon exchange in the
two limits which ultimately resolves the apparent paradox
involving the one-gluon exchange contribution to the baryon mass
discussed above.  Moreover, we show more generally that the
contribution to the mass from {\it all} types of multiple gluon
exchange diagrams scales as $N_c^2$.  This is what is required to
have the baryon mass scale as $N_c^2$, and thus to obtain
precisely the behavior needed for the baryon to scale as a
Skyrmion in the OLNC limit.

Similarly, we study characteristic diagrams contributing to
numerous quantities associated  with hadronic interaction and
from these diagrams abstract the $N_c$ scaling behavior.  In
particular, we consider the strength of the meson-baryon coupling
($N_c^1$), the baryon-meson scattering amplitude ($N_c^0$),
baryon-meson scattering to a two-meson final state ($N_c^{-1}$),
and the baryon-baryon coupling ($N_c^2$). These are precisely the
scaling rules one would expect if the baryon were a Skyrmion.

Given these scaling results, we argue that one can view baryons as
Skyrmions in the OLNC limit as well as in the TLNC limit. The
fundamental difference between the two cases is that any quantity
which scales as $N_c^{k}$ in the TLNC limit scales as  $N_c^{2k}$ in
the OLNC limit.

In the analysis that follows we will sometimes draw
representative Feynman diagrams.  Occasionally,  where it is
important to illustrate the color flow, we follow 't Hooft and 
use color-flow diagrams in which we draw gluons as two oppositely
directed color lines. In the TLNC limit, quarks are represented
by single fermion lines, while in the OLNC limit quarks are
represented by doubled fermion lines pointing in the same
direction, in order to reflect the fact that quarks now carry two
color indices. The double-line representation for quarks in the
OLNC limit will be used in both Feynman diagrams and color-flow
diagrams.

The central focus of this paper is on baryons.  However, the
identification of baryons as solitons in a mesonic theory requires
an understanding of the scaling rules in the meson sector
encapsulated in Eq.~(\ref{meson-scale}).  Moreover, the
elucidation of some aspects of the mesonic sector is essential for
clarifying the meson-baryon interaction. Accordingly, the next
section will sketch the derivation of the scaling rules for the meson sector. Since
these results are well known there is no need to be complete; we
only attempt to provide enough detail to elucidate the main points. Next,
we devote a short section to the discussion of a vital difference
in the color-flow in one-gluon exchanges between two quarks in the
TLNC and OLNC limits. This distinction will help resolve the
apparent paradox involving baryon mass scaling that was discussed
above. Following that section, we turn to the main focus of the
paper: the scaling properties of baryons. We consider classes of
diagrams which enable us to deduce the scaling of the baryon mass
and various aspects of interactions of baryons with other
hadrons.  Finally, there is a brief concluding section.

\section{Mesons \label{sec:meson}}
In this section we briefly review the large $N_c$ scaling of meson
interaction amplitudes in the TLNC and OLNC limits.  While the
results are well known, they are useful in what follows.
Throughout the section, we first review how the analysis works
for a given quantity in the TLNC limit, and then discuss the
analogous derivation in the OLNC limit. To streamline the
discussion, we examine simple quark loops as representatives of
the leading order class of diagrams for each quantity we
examine.  This can be done without loss of generality as
the inclusion of more complicated planar graphs clearly does not
alter the result.

In both of the TLNC and OLNC limits, meson masses have the same
scaling as quark masses, i.e., they scale as $N_c^0$.  Our
first step is to determine the $N_c$ scaling of the matrix element
for a current to create a meson.

\begin{figure}
    \centering
        \subfigure[TLNC limit]{
            \label{2quark:TLNC}
            \includegraphics[width=1.5in]{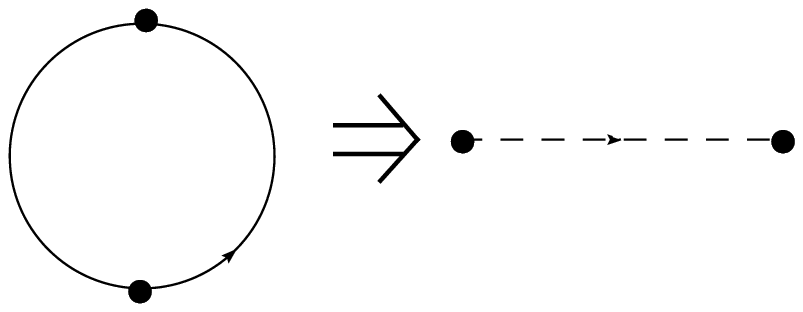}
        }
        \subfigure[OLNC limit]{
            \label{2quark:OLNC}
            \includegraphics[width=1.5in]{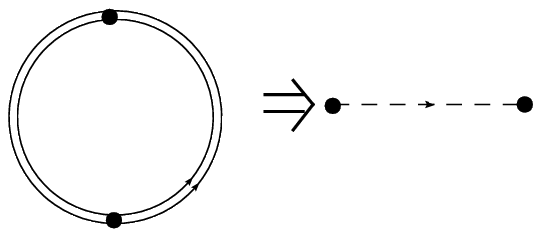}
        }
    \label{2quark}
    \caption{Quark loops with two current insertions (as representatives
    of the class of leading order diagrams for the two-point function) and their
    associated hadronic content in terms of meson propagation.}
\end{figure}

We begin with the TLNC limit.  Consider a quark loop with two
currents carrying meson quantum numbers at the edges as a
representative diagram for the two-point correlation function (Fig.
\ref{2quark:TLNC}
--- the solid dots represent the currents). There are $N_c^1$
choices of color for the quark loop, so the diagram must scale as
$N_c^1$ as a whole.   Matching the $N_c$ scaling of the diagrams
with the meson picture, one sees that the amplitude for the
current to create a meson must scale as $N_c^{1/2}$.

The analysis proceeds in an analogous manner for the OLNC limit;
the only significant difference is that there are $N_c^2$ choices
for the color loop in Fig. \ref{2quark:OLNC}, and as a result each
meson creation matrix amplitude scales as $N_c^1$ rather than
$N_c^{1/2}$.  At this point, we should note that up to constants
of proportionality, $f_\pi$ is the amplitude for the axial current
operator to create a pion from the vacuum.  The preceding
analysis shows that $f_{\pi} \sim N_c^{1/2}$ for the TLNC limit
while $f_{\pi} \sim N_c^{1}$ for the OLNC limit. This is
precisely what is needed for consistency with the Skyrme
Lagrangian as seen in Eq.~(\ref{epsscale}).

\begin{figure}
    \centering
        \subfigure[TLNC limit]{
            \label{mesonDecay:TLNC}
            \includegraphics[width=1.5in]{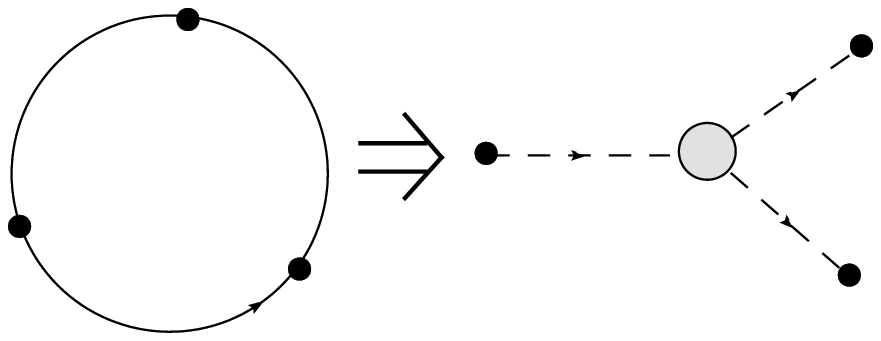}
        }
        \subfigure[OLNC limit]{
            \label{mesonDecay:OLNC}
            \includegraphics[width=1.5in]{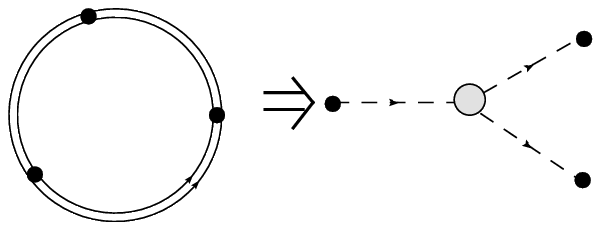}
        }
    \label{mesonDecay}
    \caption{Meson decay diagrams.
    The relationships between quark loops (as typical members of
    the class of leading order diagrams) with three current
    insertions and the hadronic-level effective diagrams are illustrated.}
\end{figure}

Now consider the $N_c$ scaling of the amplitude for the
three-meson vertex which fixes the strength of a meson decaying
into two mesons. In the TLNC limit (Fig.
\ref{mesonDecay:TLNC}), we again start with a quark loop, but this time
with three current insertions, as a representative of the class of
leading order diagrams for the three-point correlation function.
At the hadronic level this diagram represents the creation of
three mesons from the currents, with the mesons interacting via a
trilinear meson-meson-meson vertex. The diagram as a whole still
scales as $N_c^1$, but we know that each of the matrix elements
scale as $N_c^{1/2}$. This means that the trilinear
meson-meson-meson vertex must scale as $N_c^{-1/2}$.  From this we
conclude that the amplitude for meson decays scales as
$N_c^{-1/2}$, while the width scales as $N_c^{-1}$, and thus mesons are
stable at large $N_c$ in the TLNC limit.

In the OLNC limit, the main difference is again the $N_c^2$
choices of color labels for the quark loop (Fig.
\ref{mesonDecay:OLNC}).  It is not hard to see that this implies that the three meson vertex must scale as
$N_c^{-1}$ and its width therefore scales as $N_c^{-2}$; mesons
are also stable in the OLNC limit.  Note that the scaling
relation for the three-meson vertex is consistent with
Eq.~(\ref{meson-scale}).

It should be immediately clear from the preceding example
how to generalize to the case of an interaction vertex for any number of mesons.
Adding one more meson reduces the scaling by a factor of
$N_c^{-1/2}$ for the TLNC limit and by a factor of $N_c^{-1}$ for
the OLNC limit. Taken together these immediately yield Eq.
(\ref{meson-scale}).

Note that the generic replacement rule for scaling that was given in
the introduction, $N_c^{k} {\rm (TLNC}) \rightarrow N_c^{2k} {\rm
(OLNC}) $, holds throughout the meson sector.

\section{One-gluon exchange}
As noted in the introduction, in order for there to be a possibility of identifying baryons with solitons in the OLNC limit, there must be a subtle distinction between the behavior in the TLNC limit and OLNC limit.  The naive analysis of
the one-gluon exchange contribution to the baryon mass gives a result consistent with the Skyrmion for the TLNC limit and a
result apparently inconsistent for the OLNC limit.  The origin of this discrepancy can be traced to the nature of gluon exchange
between quarks in the two cases. In this section we focus on elucidating the differences in  one-gluon exchange between two quarks in the TLNC and OLNC limits.

\begin{figure}
    \centering
        \subfigure[Quarks in the fundamental representation]{
            \label{qexchange-T}
            \includegraphics[width=1.5in]{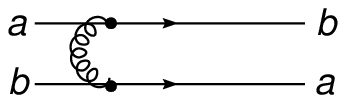}
        }
        \subfigure[Color flow for quarks in fundamental representation]{
            \label{qexchange-TF}
            \includegraphics[width=1.5in]{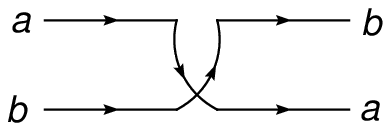}
        }
    \caption{One-gluon exchange between quarks in the fundamental representation.
    The colors $a$ and $b$ are switched by the exchange}
\end{figure}

Consider one-gluon exchange between two quarks in the TLNC limit (Fig.\ref{qexchange-T}), where the quarks are taken to be in the fundamental representation and thus are labeled by a single color.
The key point is that the effect of the gluon exchange on the
quark content is simply to switch around the color labels of the
two quarks ($a$ and $b$ in the figure), \emph {i.e.}, after the
exchange one has quarks with the same colors as before the exchange.
The reason for this is clear from the color flow diagram of
Fig.~\ref{qexchange-TF}.

\begin{figure}
    \centering
        \subfigure[Quarks in anti-symmetric representation]{
            \label{qexchange-O}
            \includegraphics[width=1.5in]{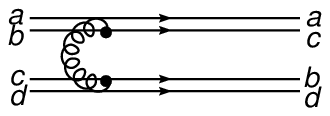}
        }
        \subfigure[Generic color flow for quarks in anti-symmetric representation]{
            \label{qexchange-OF}
            \includegraphics[width=1.5in]{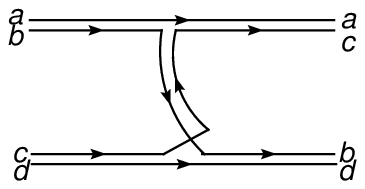}
        }
         \subfigure[Color flow for gluon exchange with no change in color labels]{
            \label{gexchange-OF-Cartan}
            \includegraphics[width=1.5in]{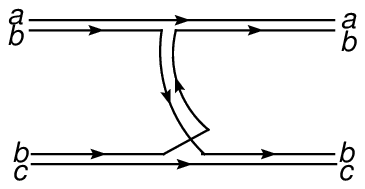}
        }
    \caption{One-gluon exchange between quarks in the anti-symmetric representation.
    The colors for the initial quarks $a b$ and $c d$ are generally, but not necessarily, distinct from the colors for the final quarks, $a c$ and $b d$.}
\end{figure}

In contrast, consider a one-gluon exchange for two quarks in the
anti-symmetric representation relevant for the OLNC limit (Fig. \ref{qexchange-O}), where each quark is labeled by \emph{two} color indices.
Note that while the total color of the state is preserved by the
interaction (one has fundamental colors $a$, $b$, $c$ and $d$
both in the initial and final state), the color labels of the individual
quarks are generally altered.  In the case illustrated in Fig. \ref{qexchange-O}, initially one has quarks of the
$a b$ and $ c d$ varieties, but after the interaction there is one quark with $a c$
and one with $b d$. The reason for this is clear from the color
flow diagram in Fig.~\ref{qexchange-OF}. Thus, unlike the situation with quarks taken to be in the fundamental representation, as in the TLNC limit, gluon exchanges typically alter the color labels of quarks taken to be in the anti-symmetric representation, as in the OLNC limit.

This fact plays a critical role in combinatoric counting at large $N_c$. If
we restrict ourselves to situations in which the colors of the
initial and final quarks must be the same (up to a permutation), then a one-gluon exchange in the OLNC limit
requires a constraint on the type of quarks which participate.  In
particular, they have to share one of their two color indices.  For
example, in the diagram in Fig. \ref{qexchange-O}, the
restriction is for $a=d$ (recalling that the labels $b a$ and $a
b$ are equivalent for an anti-symmetric representation).  This
restriction will play a key role in reducing combinatoric factors
when considering the scaling of baryonic quantities.

\begin{figure}[h]
    \centering
        \subfigure[Two-gluon exchange]{
            \label{2gexchange-O}
            \includegraphics[width=1.25in]{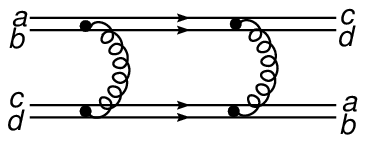}
        }
        \subfigure[Color flow for two-gluon exchange]{
            \label{2gxchange-OF}
            \includegraphics[width=1.25in]{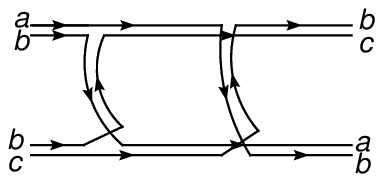}
        }
    \caption{Two-gluon exchange graphs for quarks in the anti-symmetric
representation.  The quarks can have generic initial color labels and suffer no change in final quark color labels}
\end{figure}

Two quarks can also exchange a gluon (e.g., a $b\bar{b}$, as in Fig. \ref{gexchange-OF-Cartan}) and undergo no changes or permutations in color labels.  In such situations, the gluons must be in the Cartan subalgebra --- that is, the diagonal subalgebra --- of the algebra of $SU(N_c)$ \cite{BolognesiPriv}.  It is non-trivial to show such gluons in an 't Hooft style double-line diagram, but the imposition of the tracelessness condition for the $SU(N_c)$ algebra is a $1/N_c$ suppressed effect, and thus one can simply work with gluons in the algebra of $U(N_c)$ to leading order in $1/N_c$\cite{Coleman} --- which is what we do in Fig. \ref{gexchange-OF-Cartan}.

Of course it is possible for two quarks with no shared color labels to interact with no change of color labels, but this generally requires a two-gluon exchange (Fig. \ref{2gexchange-O}).  It is clear that in a certain sense, the case of two-gluon exchanges between two quarks in the OLNC limit is analogous to one-gluon exchange in the TLNC limit.  Again, the reason this works is easily
seen in the color-flow diagram of Fig.~\ref{2gxchange-OF}.  This fact
also plays an important role in the scaling at large $N_c$, since
the two-gluon exchange diagrams have an extra factor of
$g^2 \sim N_c^{-1}$ compared to one-gluon exchange.

The preceding illustrates the central distinction between the
nature of gluon exchange between quarks in the fundamental and
anti-symmetric representations.  It makes clear that we cannot
simply copy Witten's combinatoric analysis developed for the
TLNC limit for the OLNC limit analysis.  Instead, we must modify it suitably to account for the differences in one-gluon exchanges in the two limits.   Once this is taken into account, it is straightforward to show that the baryon quantities in the
OLNC limit do in fact scale with $N_c$ in a manner consistent with
a Skyrmion.

\section{Baryons}
\subsection{Baryon mass}
In the traditional 't Hooft large $N_c$, limit baryons are
antisymmetric, color-singlet combinations of $N_c$ quarks (plus
associated gluons and those quark-antiquark contributions which
arise through ``z-graphs'' without closed quark loops \cite{CohenLeinweber}).
The quarks have a fixed mass of order $N_c^0$, yielding a
contribution to the baryon mass that scales as $N_c^1$; similarly,
the kinetic energy of the quarks is a one-body operator and its
contribution to the baryon mass also scales as $N_c^1$.  Thus, it
is is natural to assume that the baryon mass scales as $N_c^1$.
For consistency, the contributions to the baryon mass from gluon
exchange must also scale like $N_c^1$.  It is not very difficult
to verify that this is indeed the case.

\begin{figure}
    \centering
        \subfigure[TLNC limit]{
            \label{baryonMass:TLNC}
            \includegraphics[width=1.5in]{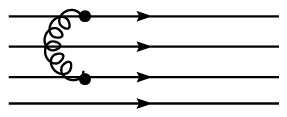}
        }
        \subfigure[OLNC limit]{
            \label{baryonMass:OLNC}
            \includegraphics[width=1.5in]{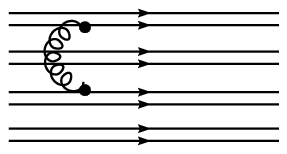}
        }
    \label{baryonMass}
    \caption{Typical one-gluon exchange diagrams that contribute to the baryon mass. The letters correspond to quark color labels before and after the gluon exchange.}
\end{figure}

Witten showed that in order to investigate gluon-exchange contributions to the baryon mass, the relevant quantities to study are the
quark-line connected diagrams (the disconnected ones arise through
exponentiation of the Hamiltonian) \cite{WittenOrig}.  Consider, as a
simple example, the one-gluon interaction between a pair of quarks
in the baryon as illustrated in Fig. \ref{baryonMass:TLNC}. As
discussed briefly above, this contribution scales as
$N_c^1$. Recall that any two quarks in a baryon can interact in this way,
since they simply exchange color indices in the interaction,
keeping the baryon a color singlet. The two quark-gluon vertices
together scale as $(N_c^{-1/2})^2=N_c^{-1}$. There are $N_c^1$
choices for the first quark involved and another $N_c^1$ choices
for the second one, giving a total combinatoric factor of
$N_c^2$. It follows that such diagrams are of order $N_c^{1}$.

Quark-line connected diagrams involving more than two quarks do
not change this conclusion because connecting an additional quark
to the diagram requires adding two new gluon vertices,
for a factor of $N_c^{-1}$, and a combinatoric factor of
$N_c^1$ from the sum over colors. As a result additional connected quarks only add
factors of $N_c^0$ to such self-interaction diagrams.  This reasoning can easily be cast into the form of an argument by induction, and a generalization of this idea will be used in the discussion of the OLNC limit below.  

Inserting additional gluons which connect to
pre-existing gluons does not alter the counting. By standard
arguments an additional gluon will at most add a closed color
loop in the sense of `t Hooft diagrams thereby adding a power of
$N_c$; this is compensated for by two coupling constants at
$N_c^{-1}$ yielding no change in the $N_c$ counting (this is the analog
of the planar diagrams from the meson case). Depending on the
topology of the diagram, additional gluons may not add a color
loop, in which case their graphs are suppressed in the $1/N_c$
expansion (these are the non-planar graphs). Additional quark
loops do not add a color loop but cost a power of $1/N_c$ from the
vertices and are thus always suppressed. From these considerations,
we see that in the TLNC limit the general gluon-exchange
contribution to the baryon mass really is of order $N_c^1$.  This is consistent with the baryon mass scaling as $N_c^1$.

In the OLNC limit the situation is somewhat more complicated.  As
shown by Bolognesi\cite{Bolognesi}, in this limit baryons are an antisymmetric
combination of $N_c (N_c-1)/2 \sim N_c^2$ quarks, each of which
now carries two color indices.  Since each quark still has a mass
and a kinetic energy of order $N_c^0$, this means that in this
limit the baryon mass should scale as $N_c^2$.  However, for this
to be true, the contribution to the mass from gluon-exchange interactions between
the quarks must also scale as $N_c^2$ in the OLNC limit.

First, consider a representative diagram of a one-gluon interaction
between two two-index quarks in an OLNC limit baryon (Fig.
\ref{baryonMass:OLNC}).  As noted in the introduction, a naive
recapitulation of the reasoning used in the TLNC case leads to
the conclusion that such diagrams scale as $N_c^3$: there are two
gluon vertices, which together scale as $N_c^{-1}$, and $~N_c^2$
choices for each of the two participating quarks, yielding a
complete diagram that scales as $N_c^{-1} N_c^{4}=N_c^3$.  This is
clearly inconsistent with the baryon mass scaling like $N_c^2$.

\begin{figure}
    \centering
        \subfigure[Forbidden interaction]{
            \label{OLNCdetail:forbidden}
            \includegraphics[width=1.5in]{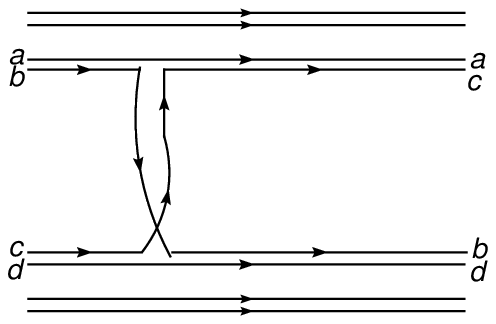}}
        \subfigure[Allowed interaction]{
            \label{OLNCdetail:allowed}
            \includegraphics[width=1.5in]{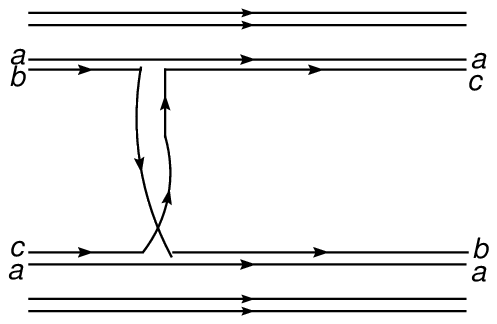}}
		
    \label{OLNCdetail}
    \caption{Not every interaction between two quarks within a baryon is allowed in the OLNC limit.}
\end{figure}

In fact, a more careful analysis shows that one-gluon interaction
diagrams in the OLNC limit scale as $N_c^2$.  The basic reason
was foreshadowed in the preceding section: as in the TLNC limit, the
interacting quarks swap a color index through the interaction,
but because each quark now carries two color indices, there are
restrictions on which quarks can interact in this way within a
baryon. For example, suppose the interacting quarks are labeled with color
indices $ab$ and $cd$. After exchanging a $b\bar{c}$ gluon, they
become labeled with the indices $ac$ and $bd$ (Fig.
\ref{OLNCdetail:forbidden}). However, since the baryon is an
antisymmetric combination of all possible two-color labeled
quarks, after such an interaction the baryon would `lose' the
$ac$ and $bd$ quarks by antisymmetry, as well as the $ab$, $cd$
quarks. Such a final state must vanish. This forces us to conclude
that quarks which do not share at least one color label cannot
interact directly via a one-gluon exchange in a baryon.

However, if two quarks \emph{do} share a color label, then a direct
interaction between them will survive. For example, two quarks
labeled $ab$ and $bc$ can interact via the exchange of an
$a\bar{c}$ gluon, and will have color labels $cb$, $ba$ after the
interaction (Fig. \ref{OLNCdetail:allowed}) --- the color labels are simply permuted.  As desired, after the interaction the baryon still consists of an antisymmetric combination of all possible two-color labeled quarks.

A one-gluon exchange within a baryon (see, \emph{e.g.}, Fig. \ref{gexchange-OF-Cartan}) that does not alter or permute any color labels and is allowed by the antisymmetry condition is also possible\cite{BolognesiPriv}.  As discussed in the preceding section, this involves gluons in the Cartan subalgebra of $SU(N_c)$.  In such diagrams, the involved quarks must share some color labels, so their $N_c$ scaling is the same as those of the other diagrams involving one-gluon exchange. 

From these considerations we see that only quarks that share a color label can
interact via a one-gluon exchange in a baryon in the OLNC limit.  Consider now the
$N_c$ scaling of a diagram of such an interaction (Fig.
\ref{baryonMass:OLNC}) in the OLNC limit.  There are two
quark-gluon vertices, for a factor of $(N_c^{-1/2})^2=N_c^{-1}$.
There are $N_c^2$ choices for the first quark involved, but only
$N_c^1$ choices for the second because it must share a color
label with the first quark, giving a combinatoric factor of
$N_c^3$.  Thus the entire diagram scales as $N_c^2$.

\begin{figure}
    \centering
        \subfigure[Feynman diagram for an interaction between two unlike quarks $ab$ and $cd$]{
            \label{OLNCmany:unlike} \includegraphics[width=1.5in]{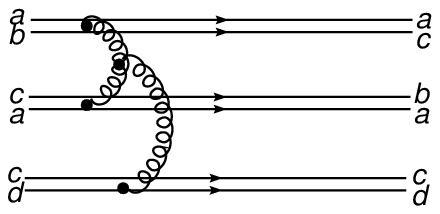}}
        \subfigure[Color flow for an interaction between two unlike quarks $ab$ and $cd$]{
            \label{OLNCmany:colorFlow} \includegraphics[width=1.5in]{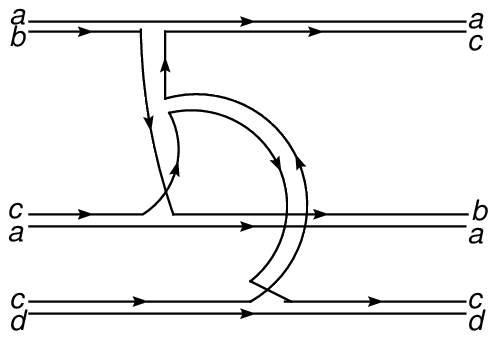}}
        \subfigure[A typical many-quark interaction]{
            \label{OLNCmany:vertices}
            \includegraphics[width=1.6in]{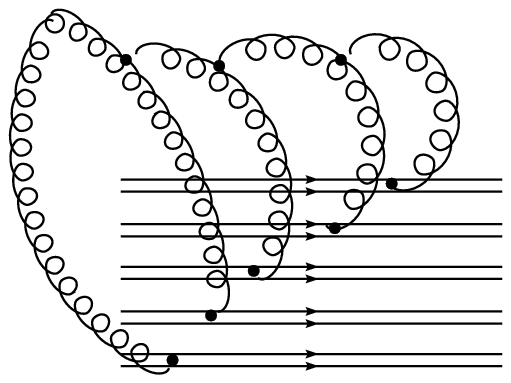}}

    \label{OLNCmany}
    \caption{Gluon-exchange interactions in a baryon between multiple quarks in the OLNC limit.}
\end{figure}

We note that two quarks in a baryon that share no color indices
\emph{can} interact with each other, but the interaction must involve more than one gluon
exchange.  If two quarks exchange two gluons directly (as in Fig. \ref{2gexchange-O}), the four gluon vertices will give a factor of $N_c^{-2}$, and the $N_c^2$ choices for the labels of each of the two quarks will result in a $N_c^2$ scaling for the interaction.  Alternatively, two quarks with unlike labels may interact via gluon exchanges with an intermediary third quark, which must share some indices with both of the unlike quarks, as in Fig. \ref{OLNCmany:unlike}.

To show that gluon exchanges contribute at most $N_c^2$ to the baryon mass, we must demonstrate that diagrams with an arbitrary number of interacting quarks within a baryon scale as $N_c^2$.  To show this, we will construct an argument by induction that shows that diagrams with $q$ interacting quarks (Fig. \ref{OLNCmany:vertices}) scale as $N_c^2$ at large $N_c$.  The argument by induction is essentially based on the idea that one can build a diagram  with $(q+1)$ interacting quarks by adding a quark to some $q$-quark diagram, and the observation that such an addition does not change the $N_c$ scaling of the diagram.

 As the base case (that is, $q=2$), we have already shown above that diagrams with two interacting quarks scale as $N_c^2$.  Next, observe that any leading-order diagram with $q+1$ interacting quarks can be constructed from some $q$-quark diagram by connecting (via one or more gluons) an additional quark.  As the inductive step, suppose that the $q$-quark diagram scales as $N_c^2$.  We can connect a new $(q+1)^{st}$ quark to the diagram in one of three ways:  either by a one-gluon connection to a quark in the $q$-quark diagram, by a one-gluon connection to a gluon in the $q$-quark diagram, or by a two-gluon connection to a quark in the $q$-quark diagram.

The first two cases above are identical as far as the topology of color flow is concerned, as an inspection of Fig. \ref{OLNCmany:colorFlow} makes clear.  Therefore, we can consider only the cases of direct quark-quark connections, without loss of generality.  Since the new quark connects via gluon exchange to a quark in the $q$-quark diagram, the situation is reduced to that of the base case of two interacting quarks.  

If only one gluon is exchanged (with a factor of $N_c^{-1}$ from the two new coupling constants), the new quark must share a color index with the quark with which it is interacting, yielding a combinatoric factor of $N_c^1$. Alternatively, if two gluons are exchanged (with a factor of $N_c^{-2}$ from the four new coupling constants), the new quark need not share any color indices with the quark with which it is interacting, yielding a combinatoric factor of $N_c^2$.  In either case, the scaling of the ($q+1$)-quark diagram is the product of the scaling of the $q$-quark diagram, $N_c^2$, and a factor of either $N_c^{-1}N_c^{1} \sim N_c^0$, or $N_c^{-2}N_c^{2} \sim N_c^0$.  Thus we see that a general ($q+1$)-quark diagram scales as $N_c^2$ in the OLNC limit.  This completes the argument by induction, and we conclude that any diagram with $q$ interacting quarks scales as $N_c^2$ at leading order.

Of course, diagrams beyond the class considered above can
contribute. For example, additional gluons can connect between the
gluons in flight yielding closed gluon loops.  However, such
additional gluon loops will not alter the $N_c$ counting.  As in
the case of the TLNC limit, adding a gluon to a diagram can at
most add a closed color loop in the  sense of an 't Hooft
diagram, adding a factor of $N_c$ which is compensated by a
$N_c^{-1}$ factor from the additional vertices.  This yields either an unchanged
$N_c$ scaling or a suppression.

Unlike the TLNC limit, closed quark loops are not suppressed in the OLNC limit. Due to
their two-index nature they behave analogously to gluons.
Depending on the topology of the diagram, quark loops can add at
most one new color loop, which is exactly compensated for by the
$N_c^{-1}$ factor due to the new vertices.  Thus, while quark loops are not
suppressed, they also do not alter the leading $N_c$ counting.

As a result of these considerations, it is apparent that
the total energy of interactions between quarks due to the
exchange of gluons is of order $N_c^2$.  Thus we see that the
baryon mass consistently scales as $N_c^2$. This is consistent
with the known scaling of the soliton mass, which is also $N_c^2$
in the OLNC limit.

\subsection{Scattering}
Our goal in this subsection is to show that the scaling rules for
scattering amplitudes and coupling constants between baryons and
mesons in the OLNC limit work analogously to the parallel
quantities in the TLNC limit with the standard substitution
$N_c^{k} \rightarrow N_c^{2k}$ required for the consistency of the
Skyrmion picture.

\begin{figure}
    \centering
        \subfigure[TLNC limit]{
        \label{baryonSpits:TLNC}
            \includegraphics[width=1.5in]{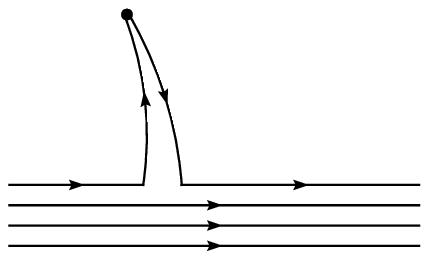}}
        \subfigure[OLNC limit]{
            \label{baryonSpits:OLNC}
            \includegraphics[width=1.5in]{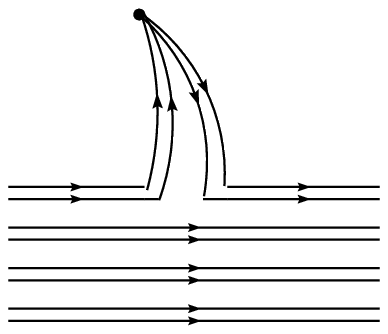}}
        \label{baryonSpits}
        \caption{Representative diagrams for the meson-baryon coupling.}
\end{figure}

We begin with an examination of the  baryon-meson vertex. First,
consider typical diagrams representing a baryon emitting a meson
(Figs. \ref{baryonSpits:TLNC}, \ref{baryonSpits:OLNC}) in the
TLNC and OLNC limits.  The dot represents a current with the
quantum numbers of some meson. One can add to these ``skeletons''
various gluon insertions (and quark loops for the case of the
OLNC limit) without altering the basic $N_c$ counting rules. The
amplitudes for coupling to a meson, as opposed to the current
itself, will be controlled by the amplitude for the creation of an
extra meson, which as shown in Sec.~\ref{sec:meson} scales like
$N_c^{-1/2}$ and $N_c^{-1}$ in the TLNC and OLNC limits
respectively.  Thus, one generically expects that the meson-baryon
coupling constant will scale as $N_c^{1/2}$ (TLNC limit) or
$N_c^{1}$ (OLNC limit).  This is consistent with the
identification of a baryon as a soliton: the soliton-meson
coupling generically scales as $1/g$.  Thus, as is expected, the
scaling matches provided $g \sim N_c^{-1/2}$ (TLNC limit) or $g
\sim N_c^{-1}$ (OLNC limit).

\begin{figure}
    \centering
        \subfigure[TLNC limit, Feynman diagram]{
        \label{meson1Baryon:TLNC}
            \includegraphics[width=1.5in]{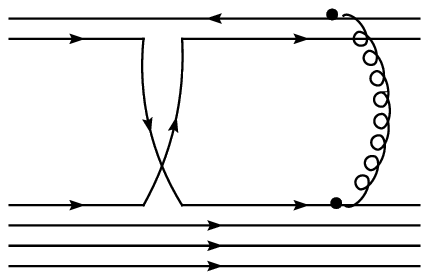}}
        \subfigure[OLNC limit, Feynman diagram]{
            \label{meson1Baryon:OLNC}
            \includegraphics[width=1.5in]{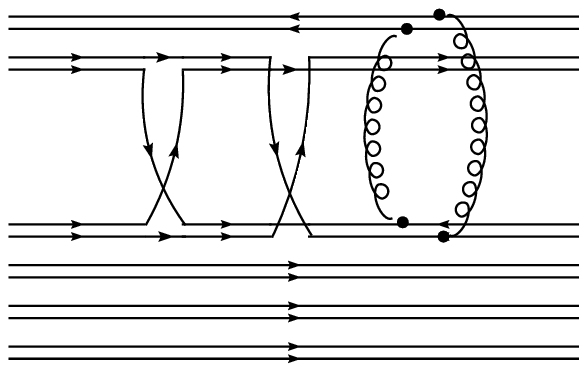}}
        \label{meson1Baryon}
        \caption{Representative diagrams contributing to meson-baryon scattering.}
\end{figure}

Next consider meson-baryon scattering.   First consider  the TLNC
limit.  A characteristic diagram contributing to the process
(Fig. \ref{meson1Baryon:TLNC}) is the exchange of a quark
between the baryon and the meson; following this exchange there
must be a gluon exchange to keep the baryon and meson separately
color singlets.  In such a graph, there are two gluon vertices
(for a suppression of $1/N_c$), and a combinatoric factor of
$N_c$ since $N_c$ different quarks in the baryon can participate
in the exchange. As a result, typical diagrams for baryon-meson
scattering scale as $(N_c^{-1/2})^2 N_c^1 = N_c^0$. This result
is consistent with meson-soliton scattering with the standard
identification since the meson-scattering amplitude is independent of $g$ at large $g$.

Now consider an analogous diagram in the OLNC limit (Fig.
\ref{meson1Baryon:OLNC}).  As before, the interaction takes the
form of a quark exchange.  However, since the quarks now carry
two color indices, there must be at least two gluons exchanged in
order to keep the baryon and meson separately color singlets .
As a result, there are four gluon vertices in a representative
diagram, contributing a total of $(N_c^{-1/2})^4=N_c^{-2}$, and a
combinatorial factor of $N_c^2$ due to the sum over the possible
color labels for the quark in the baryon participating in the
interaction.  The complete diagram thus scales as $N_c^0$, just
as before.  Again this is consistent with a soliton description.

\begin{figure}
    \centering
        \subfigure[TLNC limit, Feynman diagram]{
        \label{meson3Baryon:TLNC}
            \includegraphics[width=1.5in]{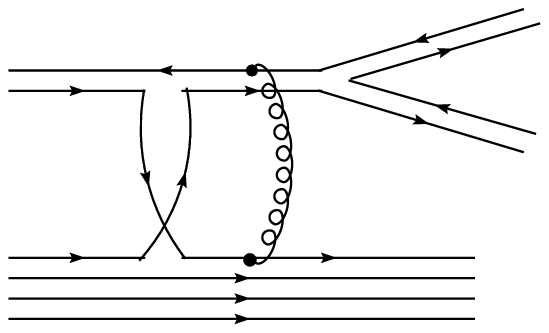}}
        \subfigure[OLNC limit, Feynman diagram]{
            \label{meson3Baryon:OLNC}
            \includegraphics[width=1.5in]{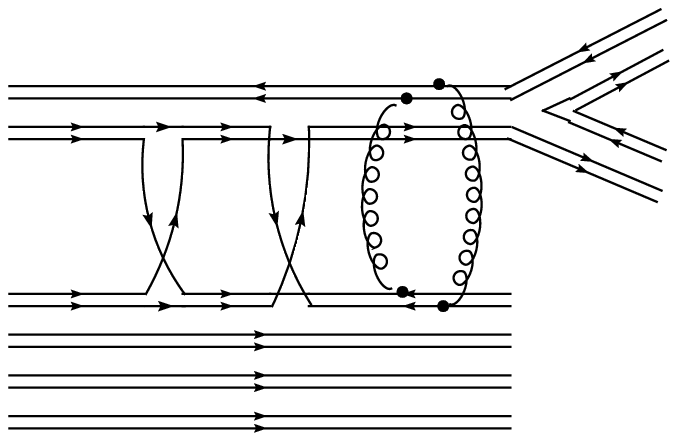}}
        \label{meson3Baryon}
        \caption{Representative diagrams for meson-baryon scattering with a two-meson final state.}
\end{figure}

Next consider a scattering process in which an incident meson on a
baryon yields a final state with two mesons. We first review the
situation in the TLNC limit (Fig. \ref{meson3Baryon:TLNC}). An
incoming meson interacts with a baryon as in the meson-baryon case
(by a quark exchange plus a  gluon interaction), and then decays
into two outgoing mesons.  The first part of this interaction,
involving the baryon, scales as $N_c^0$ as argued above.  For the
second part, involving a meson decay, one may recall from above (in Sec. \ref{sec:meson})
that the amplitude for such a process scales like $N_c^{-1/2}$.
Thus the complete diagram scales as $N_c^{-1/2}$ in the TLNC
limit.  As before, this is consistent with a soliton description, in
which such a process scales with the generic meson coupling constant, $g$, as $g^1 \sim N_c^{-1/2}$ in the TLNC limit.

For an analogous diagram for the OLNC limit, we claim
that diagrams that show scattering with an initial meson on a baryon
yielding  a two-meson final state scale as $N_c^{-1}$ (Fig.
\ref{meson3Baryon:OLNC}). As before, the baryon-meson interaction
scales as $(N_c^{-1/2})^4 N_c^2=N_c^0$. Recalling the result for
meson decays in the OLNC limit, we see that the meson decay part of the
diagram now scales as $N_c^{-1}$. Thus the full diagram scales as
$N_c^{-1}$.  Again, this is consistent with a soliton description,
since in the OLNC limit the generic meson coupling constant $g$ scales as $N_c^{-1}$.

\begin{figure}
    \centering
        \subfigure[TLNC limit]{
        \label{baryonBaryon:TLNC}
            \includegraphics[width=1.5in]{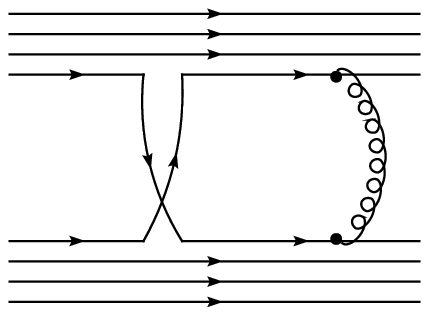}}
        \subfigure[OLNC limit]{
            \label{baryonBaryon:OLNC}
            \includegraphics[width=1.5in]{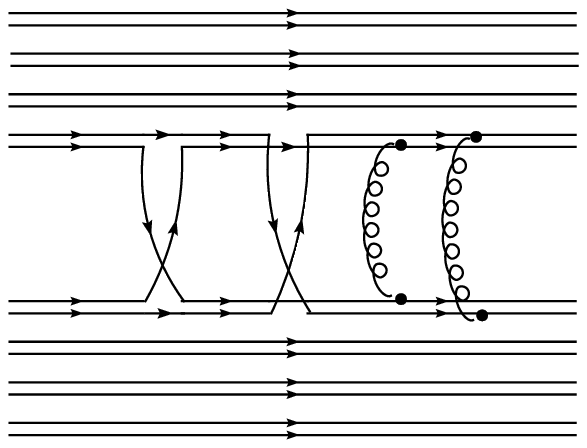}}
        \label{baryonBaryon}
        \caption{Representative baryon-baryon interaction diagrams. }
\end{figure}

Finally we consider baryon-baryon scattering.  As noted by
Witten, the kinematics of this situation are peculiar. Since the
mass grows with $N_c$, the description of baryon-baryon scattering
at large $N_c$ ultimately turns out to be smooth in the limit
where the mass and momentum go to infinity at large $N_c$, in
such a manner that the velocity $p/M$ remains
fixed\cite{WittenOrig}.  It should be noted that it is precisely in
this limit that generic soliton models have well-defined
scattering amplitudes as $g \rightarrow 0$.  Secondly, the
natural way to describe the situation is through the overall
strength of interactions during the process --- essentially the
non-relativistic potential between the baryons
\cite{KaplanManohar, KaplanSavage} which will ultimately be seen
to be strong --- and not through the scattering amplitude. As noted by
Witten, if the energy of interaction is comparable to the
incident kinetic energy, the two can play off each other in a
smooth way. Thus, the quark-line connected diagrams between
baryons should be interpreted in terms of the potential.  The
iteration of these between propagating individual baryons gives
the full amplitude.

In the TLNC limit, a representative diagram for this is Fig.
\ref{baryonBaryon:TLNC}. The two baryons exchange a quark and
also a gluon in order to stay as individual color singlets.  There
are $N_c$ choices for each of the two quarks to participate in the
interaction, and two gluon vertices, meaning the entire diagram
scales as $N_c^2 (N_c^{-1/2})^2=N_c^1$.  This means that the
baryon-baryon potential is of the same scale as the baryon mass
and kinetic energy (with fixed velocity).

The situation in the OLNC limit (Fig. \ref{baryonBaryon:OLNC}) is
analogous, but because each quark now carries two color
labels,  two gluons must be exchanged to keep the baryons color
singlets.  The $N_c$ scaling of such scattering diagrams is
simply $(N_c^2)^2 (N_c^{-1/2})^4 = N_c^2$.  Again, we see that
the baryon-baryon potential has the same scaling as the baryon
kinetic energy. Again, this result is fully consistent with
soliton-soliton scattering, where the energy of interaction
between the solitons during the collision is of order $1/g^2$.

\section{Conclusions}

Using the results of \cite {Bolognesi} for the allowed
representations for baryons in the OLNC limit, we have shown that
the baryon mass scales as $N_c^2$; this holds even when
quark-quark interactions through gluon exchange are taken into
account.  In doing so, we have resolved the apparent paradox that
a naive generalization of Witten's counting for one-gluon
exchange appears to scale as $N_c^3$. More generally, we have
shown how to generalize  Witten's analysis of baryon and meson
behavior to the OLNC limit, and demonstrated that the replacement
rule $N_c^k \rightarrow N_c^{2k}$ is justified for all of the
representative diagrams.

From this analysis, one can conclude that all of the arguments
for identifying baryons with solitons (such as the Skyrmion) in
the 't Hooft large $N_c$ limit apply to baryons in the orientifold
large $N_c$ limit.  In general, to use the original Skyrme model to
model baryons in the orientifold large $N_c$ limit, one must
simply scale $f_{\pi}^2 \sim N_c^2$ and $\epsilon^2 \sim N_c^2$
from Eq. ($\ref{SK})$, and also scale the WZW coefficient as $n \sim
N_c^2$ in accordance with the replacement rule above.  This will
ensure that all of the generic hadronic scaling rules behave correctly.

We note, however, that {\it the} Skyrme model (i.e., Skyrme's original model) is {\it not} justified in
large $N_c$ limit. What is presumably justified is {\it a} Skyrme-type
soliton model with an arbitrary number of fields and arbitrarily
complex interactions.  The justification for such a model based
on generic scaling rules for QCD in the OLNC limit is essentially
the same as in the TLNC limit.

We should also note that of course the Skyrmion encodes more than
just the generic scaling rules, as it also encodes large $N_c$
scaling rules associated with spin and flavor.  Relations between
observables sensitive to spin and flavor in Skyrme-type modes but
independent of the dynamical details of the particular model were
noted early on by Adkins and Nappi \cite{AdkinsNappi}.
Subsequently, it was noted first by Gervais and
Sakita\cite{GervaisSakita} and then developed in considerable
detail by Dashen and Manohar \cite{DashenManohar} and Dashen,
Jenkins and Manohar \cite{DashenJenkinsManohar} that such
relations stem from large $N_c$ consistency conditions.  Since
the key to this derivation is the fact that the pion-nucleon
coupling constant grows with $N_c$ while the pion-nucleon
scattering amplitude does not, one expects that all of these
relations will go through without essential change from the TLNC
limit to the OLNC limit, again supporting the Skyrmion picture.

Although it is clear that a Skyrme-type model is capable of
describing both limits (with the parameters having a different
scaling with $N_c$ as one goes from one to the other), there
clearly are distinctions between baryons in the TLNC limit and
the OLNC limit stemming from the non-suppression of quark loops
in the OLNC limit.  How these distinctions may be manifest in
Skyrme-type models will be the subject of a future publication.

The support of the U.S. Department of Energy through grant DOE-ER-40762-366 is gratefully
acknowledged.  The authors also gratefully thank Adi Armoni, Stefano Bolognesi, Rich Lebed, Misha Shifman and Gabriele Veneziano for useful discussions.

\end{document}